# Hybrid Secure Routing in Mobile Ad-hoc Networks (MANETSs)

SOUNDES OUMAIMA BOUFAIDA [1], ABDEMADJID BENMACHICHE [1], MAJDA MAATALLAH [1] and CHAOUKI CHEMAM [1]

[1] Department of Computer Science, LIMA Laboratory, Chadli Bendjedid University, El-Tarf, PB 73, 36000, Algeria

*E-mails*

*s.boufaida@univ-eltarf.dz; benmachiche-abdelmadjid@univ-eltarf.dz;*

*maatallah-majda@univ-eltarf.dz; chemam-chaouki@univ-eltarf.dz.*

**Abstract**

Because wireless communication is dynamic and has inherent defects, routing algorithms are crucial in the quickly evolving field of mobile ad hoc networks, or MANETs This study looks at the many security problems that MANETs encounter. These problems, which pose major risks to network performance, include flooding, sinkholes, and black hole assaults to address these challenges. We introduce the Hybrid Secure Routing Protocol (HSRP), which enhances the security and robustness of routing operations by fusing trust-based tactics with cryptographic approaches. HSRP combines the strengths of both proactive and reactive routing strategies, enabling it to adapt dynamically to evolving network conditions while protecting against malicious activities. We use extensive simulations with Network Simulator (NS-2) and a thorough review of the literature to assess HSRP's performance under different attack scenarios. The results show that, in comparison to traditional protocols, HSRP increases throughput and decreases latency, hence improving routing efficiency while simultaneously bolstering data transfer security. With uses in vital domains including military operations and disaster response, this study provides a scalable and workable approach for safe routing in MANETs. The findings highlight how crucial it is to include cutting-edge security features in routing protocol design to guarantee the dependability and integrity of MANETs in practical situations.

## 1. Introduction

MANETs are characterized by self-organization, without the use of any centralized control or administration. As these networks are infrastructure less, the nodes in the network act as routers. The routes in the network are discovered by the nodes using different routing protocols. MANET protocol diversity, and user mobility, are responsible for the variation in security attacks on routing protocols. Routing protocols of MANETs lack security features. An insecure routing protocol can wreak havoc on network security, making it easy for adversaries to exploit it for malicious purposes while remaining undetected. Sometimes these protocols start misbehaving because of security attacks and distorting various services of the environment. Security problems associated with MANETs and solutions to achieve more reliable routing are the focus of these studies [1].

A distributed denial of service (DDos) attack is a severe threat. Like denial-of-service attacks on local area networks, internet application servers can also be attacked. MANET can be corrupted by overwhelming radio frequency (RF) signal. Nodes in the vicinity of such attack would be unable to communicate. Much larger MANETs would be difficult to corrupt with RF jamming signal, but smaller MANETs can be congested very easily [2]. In the case of a jamming attack, nodes cannot communicate due to RF signal interference. The network can be most vulnerable if only network control packets are jammed as nodes are most susceptible to this.

### 1.1. Background of MANETs

MANET consists of mobile nodes forming a mesh network without fixed base stations. Generally, MANETs are infrastructure less, self-configured, multi-hop wireless mobile networks formed in dynamic topology with mobile nodes. Depending on the applications, size, etc. some nodes can run on batteries, which limits the energy at respective nodes [1]. Speedy deployment at any desired location is an advantage for MANET that is enabled by portable laptops, handheld devices, etc. setting up a network without the need for infrastructure is the goal of such networks. Sending and receiving information can take place in MANETS, with packets routed by intermediate nodes that replicate the functionality of routers in a wireless network by forwarding packets based on information gathered about transmitter nodes. In wireless networks, radio wave transmission takes place for communication. Radio transmission has properties like reflection, diffraction, delay spread and scattering that lead to problems of multipath propagation. High-frequency smaller wavelength radio waves in an attribute frequency range of AKHz to GKHz are also heard among the MANET problems where pulses generated by antenna become wide and increase bit error rate.

Wireless mobile networks have recently gained significant interest due to the pervasive use of wireless technologies, especially wireless LANs [3]. Wireless technologies enable economic and easy installation and are suitable for mobile users. Wireless mobile networks can be broadly classified into two categories: infrastructure based wireless networks and MANETS. A wireless communication network with a base station called access point or land station, is termed as infrastructure based wireless network. However, for an ad hoc network no predefined fixed infrastructure exists. A mobile ad hoc network is a collection of autonomous nodes which can move in and out of the network freely. Hence, the network topology changes dynamically in a non-predictable manner. In addition to these complexities, wireless communication is susceptible to noise, interference, multipath propagation, fading, etc. All these factors make the design of MANETs challenging.

## 1.2. Importance of Secure Routing in MANETs

In the last decade, a great deal of research has been conducted on wireless networks, inspired in part of their potential applications. Wireless networks are networks that interconnect mobile devices without a wiring infrastructure. Current standards support devices with short range of communication from 10 m up to 100 m, such as Bluetooth and IEEE 802.11. MANET needs a security solution to offer protection against various types of attacks. In MANET, a mobile node(s) can join or leave the network at any time without any prior notice which makes consistent and reliable routing very difficult. These networks are more vulnerable to various types of attacks because mobile nodes can work on the network with less physical protection. MANETs consists of group of mobile nodes that communicate with each other without any pre-existing fixed infrastructure. A MANET is an autonomous system of mobile hosts connected by wireless links. Each node in MANET acts as both routers and hosts which can move in arbitrary fashion. But due to such nature of networks, one hop or multi-hop wireless communication among the nodes is exposed to various security challenges and attacks involving different threats and at different scales. Either by machinery or human operators, in any situation (natural disasters, emergency operations, war fields), adversary users can attack all or one selected node(s) by hindrance or by benign nodes to capture and disclose some private information.

Routing protocols are necessary to discover paths through an ad-hoc network between pairs of wireless devices. The alternative paths are less loaded, or higher data rate. A proper routing protocol selects between several paths for which the connections were originally set up. However, to meet the actors' requirements, the protocol must also keep the paths selected up to date. With a continually changing network topology, errors can occur in the paths. For detecting networking failures there is generally periodic monitoring of the paths as well. Due to hostile conditions, the nodes in MANET face attacks that can do the routing information modification or dropping of any messages. MANET provides perfect environment for intruder, malicious users which can corrupt the entire routing table due to the absence of base station. As this network does not have any centralized or fixed infrastructure, a mobile node(s) can join or leave the network at any time which may cause hindrance in its connectivity. So, there is a need for a secure routing protocol that can offer reliable and stable routing.

## 1.3. Overview of Trust-Based and Cryptographic Approaches

A routing protocol can be defined as a set of rules by which a node/host obtains a route to all other nodes within the same logical network. There are different protocols for different types of networks i.e. wired, wireless, ad-hoc and peer-to-peer networks. Routing protocols for Wireless network: AODV, DSR, SRP, DSDV, WRP. Trust Issues in Routing is an effort to study the security problems associated with MANETS and the methods or solutions through which routing can be made more reliable [1]. There is a certain class of issues that occur due to Ad-hoc Network Characteristics.

Trust-based approaches use social concepts like reputation to contribute toward securing networks. Reputation systems use past recommendations or individual observation and assess trust bounds of nodes based on those observed interactions and/or recommendations, to either allow or deny forwarding packets via nodes. Cryptographic approaches leverage over-sharing trusted authorities and public keys to share security parameters, which are subsequently used to create security associations. Given that such assumptions are not viable in MANETs, cryptographic techniques mainly rely on digital signatures or certificates to provide security against tampering and/or forgery and replay and/ or masquerading attacks. Unfortunately, such techniques are susceptible to internal attacks by compromised nodes that violate terms of use of the utility arguably for individual pecuniary profit. External attacks are also possible by attempts to forge signatures or assimilate false certificates therefore either being able to fabricate messages or being tamable to misuse [4].

## 2. Trust-Based Routing in MANETs

MANET is a self-configuring Infrastructure less network of mobile devices connected by wireless links.

Each node in the network may act as a host and as a router. It works on cooperative basis to perform network functions. Thus, networking protocols designed for wired networks are not directly suitable for MANETs. As the nodes are mobile, they are free to move in any direction and there is a frequent change in topology. These constant changes create challenges such as availability of complete and correct node information in a timely manner. Trust, security and quality of service are essential criteria in selecting a route to other nodes for transferring data among them [1]. This project studies different trust-based routing protocols in MANETs. A well analyzed trust model is implemented in preventive as well as reactive approach to enhance the performance of networks.

## 2.1. Concepts and Principles

With the advancement of computing and mobile devices, wireless adhoc networks, or more precisely, MANETs, have become an area of tremendous interest and research. Nodes are connected to each other dynamically in an arbitrary fashion, but communication takes place over multiple hops due to the limited transmission range of individual nodes [1]. Nodes can be added to or deleted from the network at any time, and the network topology keeps changing randomly and frequently. These types of networks have become essential for various military applications, such as battlefield surveillance, combat operations, search-and-rescue operations, etc. Furthermore, since all the nodes are free to enter and leave, there is no fixed topology which results in the dynamic and infrastructure-less nature of a mobile ad hoc network and a pose for secure routing protocols and the development of active security [5].

Secure routing Protocols are essentially responsibility for connecting mobile nodes in a mobile ad hoc network (MANET) securely. Mostly nodes in MANET are mobile and each node acts as both a host and a router. Furthermore, there is no fixed topology, and nodes are free to enter and leave. So, nodes are more introduced to the vulnerability of being misused by attacks. All routing packets are forwarded between nodes in a free manner. This creates a chance to perform various security attacks during routing. The Secure/Trusted Routing algorithm will try to increase the security of routing in MANETS. The purpose of daily computing is to securely connect all users to an unknown network. In the context of MANETs and peer-to-peer systems, the users are mobile devices that dynamically join and leave the network and act as servers and clients. Meanwhile, these devices cannot be fully trusted and can act cooperatively or maliciously.

## 2.2. Trust Metrics and Models

In trust-based routing, nodes should evaluate the level of trust to communicate with neighboring links. The most common approaches to evaluate trust metrics are quantitative metrics and qualitative metrics [6]. In MANET, links between nodes are created dynamically; users should build a trustable relationship before establishing a communication link. A quantitative trust metric is defined by a node that uses communication requests and responses to calculate the degree of trust in the connection link. Trusting nodes are classified as good, neutral, and bad based on numeric values, where $0 <=$ trust $< 0.5$ is bad, $0.5 <=$ trust $< 0.8$ is neutral, and $0.8 <=$ trust $<= 1$ has a good trust level. The trust level of a communication link is calculated using the number of direct transactions of one node and its neighbors (engagement process), assessments provided by the other node (reputation process), and observations made by the intermediary nodes (recommendation process). On the other side, each node maintains a trust table that keeps track of how other nodes are assessed with reference to the communication transactions in which it was part either as a source node or as a destination node.

A positive or negative weight is assigned based on their behavior. On the other side, the qualitative metrics may be proposed as a node assigns a qualitative label to the communication links: link strong if no bypass is detected, link normal if there are some bypasses but are less than a threshold, and link weak if there are many bypasses. With qualitative trust metrics, it can be easier to justify a certain behavior to outside observers because it is less sensitive to small variations [7]. Much of the original data supporting the use of behavioral modeling approaches in education[8] was based on learners and can similarly be adapted to identify abnormal behaviors of nodes based on their communications, using these same approaches [9] [10].

## 3. Cryptographic Techniques in MANETs

### 3.1. Digital Signatures

Digital signatures are based on asymmetric key cryptography, which uses two keys: the private key for signing and the public key for verification. Using digital signatures, the sender can generate a signature for the data using its private key. The receiver can verify the authenticity of the data and signature using the sender's public key.

Digital signatures have become a fundamental tool in achieving security properties such as data authenticity and non-repudiation. However, due to resource constraints and special characteristics, the implementation of digital signatures in MANETs needs to be adapted. To withstand forgery attacks on mobile nodes and prevent malicious activities such as altering the source of a message or replaying an old message, digital signatures need to be personalized for each node in a MANET [1]. Thus, asymmetric key cryptographic parameters for digital signatures are generated apart from the regular parameters of a mobile node. Long static secret keys (private keys) are safely stored in a tamper-resistant smart card. The public keys are stored in a public directory and maintained by a trusted key generation center.

### 3.2. Encryption Algorithms

In the next section, MANETs are described where a proactive encryption algorithm is suggested. There are two networks involved: the first is the infrastructure network, and the second is the anehdrouter network with nodes on the roadside using MANET and DSRC protocols. In the anehdrouter network, each node is the router. Mobile devices use the device's Wi-Fi or Bluetooth to connect to a node. Then the node chooses a path and routes packets until it reaches the destination node. Therefore, there is a possibility of being exposed to external attacks. Tiny encryption algorithm (TEA) is proposed to encrypt data. TEA belongs to the class of block ciphers with a block size of 64 bits, utilizing a 128-bit key with a simple structure. It consists of 32 rounds of processing where one round of TEA accepts a plaintext consisting of two 32-bit halves. A compression function that works iteratively applies basic operations, including addition, XOR, and shifts. The design of TEA is a trade-off between performance and resistance to cryptanalysis. TEA was initially designed for 32-bit processors, although more hardware-efficient implementations can be easily adapted. By using a large block cipher such as the AES, the Man-in-the-middle will be more expensive [1].

Then another block cipher is suggested which is known as two fish algorithms. Two fish were selected as one of five finalists for the advanced encryption standard (AES). It has a block size of 128 bits and a key size of up to 256 bits. It has some attractive features such as: Two fish is faster than triple DES on most platforms. It is believed to be the most efficient AES finalist for software implementations on many computing platforms. Claiming that it has a good performance in hardware implementations as well. It is known to be very secure to all known attacks, and the design methodology of two fish was specifically intended to be practically attack resistant. No attacks against two fish more efficient than exhaustive key search are known. Two fish use the same core structure as blowfish: a 16-round Feistel network using two independent subkeys for each round. The structure of two fish is very different from that of family of block ciphers, including DES, AES and RC529 which rely entirely on lookup tables for diffusion.

## 4. Challenges in Secure Routing in MANETs

MANETs permit nodes to move freely and can quickly perform communication in the absence of fixed infrastructure. These networks find many applications in rescue operations, military deployments, commercial product campaigns, and environments with no fixed infrastructure like meetings, conferences, and classrooms [11]. Each node in a MANET employs wireless communications to reach nodes within its transmission range, while other nodes outside this range receive packets after it has been relayed by other nodes thus creating a multi-hop network. However, wireless channels are inherently error-prone because of various reasons like transmission noise, interference, mobility of nodes, and signal distortion. Topological changes take place because of node initialization or expiration, node mobility, link failures, or interference created by obstacles, causing routes to be free or unestablished. Due to this, the structure of the network continuously generates routes discovery and maintenance traffic, thus generating high overhead and processing delays [1]. Security is an essential constraint due to the increase in the number of applications for mobile ad-hoc networks. MANETs are prone to security threats as the medium is open, and therefore sensitive information can be compromised. Furthermore, the protocols, information, and bandwidth are open to outside access, where malicious users can join the network to gain confidential information or affect the routing, thus draining the resources. Thus, with a highly dynamic network topology combined with resource constraints (limited battery power, on-board memory, CPU power, information processing/forwarding capacity, and collision susceptibility) represent a major challenge for the design and implementation of secure routing protocols.

### 4.1. Dynamic Network Topology

The mobile nodes in MANETs compose the network's dynamic topology through the movements of nodes, which leads to changes in links over time. Several factors contribute to the dynamic topology in a MANET, including geographic location, node mobility variation, network environment, and life period. Each factor is elaborated upon here.

Geographic Location: The geographic location of a node affects its movement. For instance, in a battlefield MANET, nodes comprising mechanisms like cameras, radars, and air-attack systems located nearby each other

(i.e., close in their geographic location) are likely to move in similar geographic terrain and thereby lead to similar movements. As a result, these nodes form a dynamic topology.

Node Mobility Variation: The variation in node mobility influences dynamic topology. The MANET scenarios having low mobility or high mobility create similar problems in maintaining the topology. For instance, in a network with low node mobility, the routes are stable and remain intact, while nodes move from one segment to another rapidly; often, rerouting does not occur within limits of the time-to-live (TTL) that specifies the packet lifetime.

Network Environment: Network environment determines the node movement pattern. For instance, in a scenario of rural/guided movement, nodes travel in a uniquely directed path and access the network subsequently or disband it (i.e., there is no further opportunity for communication). This leads to formation of two topologies: (1) initial topology prior to movement, and (2) final topology after movement.

Life Period: The life period of all nodes determines the node movement pattern. For instance, when all the nodes are equally likely to die and thereby form a dynamic topology and network segment, there is an abrupt increase in disconnected nodes. In addition, in a scenario where all nodes are assigned to the lifetime (TTL) such that they randomly die after the period, there is an abrupt shift from connected to disconnected topology.

## 4.2. Resource Constraints

The design and deployment of routing protocols in MANETs are complicated by resource constraints. The nodes are battery powered, resulting in limited energy resources that directly affect the lifetime of the network. In a MANET, every node has a wireless link of a limited bandwidth, which when not utilized correctly may cause congestion and packet dropping in the transmission. The resource constraints are not limited to energy and bandwidth. There are several protocols like IPsec that require high nodes processing capabilities that many of the mobile nodes cannot meet [1].

Considering the resource constraints, the secure routing protocols must be designed, which should consume less energy and bandwidth in the secure route establishment and maintenance. The protocols not only have to look for the most energy-efficient route, and less delay routes, but they also must look for routes with less overhead packet size for their security schemes.

## 5. Hybrid Secure Routing Protocols

Hybrid approaches that take advantage of both trust-based and cryptographic approaches provide the best of both worlds: the flexibility, scalability, fault-tolerance, and low overhead of the trust-based approach coupled with the security assurances provided by public key infrastructures and cryptographic mechanisms. Recent works in this area are completely classified, compared, and discussed in detail. The trust concept is based on subjective evaluation of other agents and is getting rapidly popular because it provides a decentralized way to ensure security where the existing solutions fall short [5]. However, trust-based solutions cannot completely substitute existing security mechanisms because of their limitations. On the one hand, cryptographic mechanisms provide a high level of security with strong characters; but, on the other hand, they are too rigid, inflexible, and costly for many scenarios. During certain phases of trust evaluation, such as in the proactive stage, these solutions require costly signature verification for all received messages which, on very low power nodes, might be a problem. To the authors, there is the initial work around HSRPs for ad hoc networks. Numerous trust-based protocols have been proposed in the last years; most of them rely on accurate modeling of the network to assess node trust values [1]. Although accurate, trust models highly depend on scenarios for which they were designed. When these models are used in different scenarios or when the modeled systems change, the correctness of the model will decrease, and it might provide completely wrong predictions or estimations. All these predictions or estimations should be handled with care; thus, trust-based systems should be considered as crude methods that have only relative and subjective effectiveness. For example, in extreme situations, such as islanded systems (i.e., where the network nodes are temporarily separated from the global networks or from the main trusted nodes which are used for trust evaluation), inaccurate predictions are possible. In alternative scenarios such as complete cooperation among nodes, all those assumptions become invalid, and under such a cooperation assumption, nodes trust each other completely and there is no need for trust management. Further, in trust-based solutions, trust values reflect the past behavior of nodes, which might not be appropriate for the current state.

## 5.1. Integration of Trust-Based and Cryptographic Approaches

The integration of trust-based and cryptographic approaches within HSRPs is examined. The regulatory standards and governance structures help to ensure cooperation and reduce the threat of attacks. Working in the trust approach gives more confidence in a route, thus, cracking down the information pass-through easily. In contrast, the cryptographic approach deals with the security margins and enhances the level of encryption. The routing protocols under the existing hybrid approach balance the trust and cryptographic schemes. Hence, the hybrids work in complementary nature of approaches to fortify the security in MANETs [12].

The collaboration of cryptographic and trust-based approaches gives not only the strength but also reduces the drawbacks, as shown in Table 6. The cryptographic approach helps in the exchange of secret keys and thus eliminates the trust creation in each communication. The advantage of these hybrid protocols is to lessen the penalty cost during the time of attacked nodes which do not cooperate and misbehave. Moreover, the prevention of attacks is handled at the routing community level, i.e., the detection of misbehaving nodes during the route construction phase itself. Thus, the likelihood of such nodes coming into the communication area is minimized [13].

### 5.2. Advantages of Hybrid Protocols

HSRPs exploit the advantages of both trust-based secure and cryptographic secure routing protocols. The trust-based secure routing protocols are itself a good option for securing routing under partially high threats as they are well suited for such environments and give high performance. The cryptographic secure routing protocols, on the other hand, are better options in case of high threat environment. However, cryptographic secure protocols still have some robustness. Therefore, the combination of trust-based secure and cryptographic secure routing protocols will offer better performance as compared to trust-based secure routing protocols in high security threat environment and improve robustness as compared to cryptographic secure routing protocols in moderate security threat environment [12].

Trust-based secure routing protocols are an appropriate option for the environment when either the nodes will have trust in some nodes and those nodes are suspected to be malicious and at the same time it is needed to secure the routing protocol from them. While the cryptographic secure protocols offer higher security features, they are less resilient and adaptive as compared to trust-based protocols. Since both the techniques offer significant advantages when used in their appropriate environments, the combination of trust-based and cryptographic secure routing will produce an HSRP which will offer good resilience and adaptiveness due to the trust-based protocols and at the same time will have good security features due to the cryptographic secure routing protocols.

Classification optimization techniques [14] and semantic extraction techniques [15] help drive more accurate routing decisions, as well as improve service discovery and context-aware communication in MANETs.

## 6. Research Methodology
### 6.1. Literature Review

The literature review was conducted to examine existing knowledge and research on MANETs security and hybrid routing protocols. A search of scholarly articles and conference papers was carried out using Google Scholar, Springer Link, IEEE Xplore, and Wiley Online Library. The following keywords were used: "Mobile ad-hoc networks," "MANETS security," "Security in MANETS," "Hybrid routing protocols," "Hybrid secure routing in MANETS," and "MANETS routing with security". Articles published between 2005 and 2025 were examined, as studies conducted after 2020 are currently limited to pseudocode or simulation design only. As a result of the search, a total of 50 scholarly articles and conference papers were selected for full-text reading based on their title and abstract. This process resulted in 34 scholarly articles and conference papers that provided detailed information on the research topic and were selected for this literature review.

The referenced publications include sources related to fields like E-Learning, smart cities, medical imaging, and financial services, to name just a few; however, all the various AI models, optimizations, and security frameworks discussed and developed in these works can also be applied directly to MANETs for a variety of reasons. The commonalities that support the application of these various works to MANETs are decentralization, dynamic behaviour, limited labelled data, and susceptibility to attack various malicious sources.

A recent trend in ad-hoc network routing is the reactive on-demand philosophy where routes are established only when required. Stable routing, security and power efficiency are the major concerns in this field. This paper is an effort to study security problems associated with MANETS and solutions to achieve more reliable routing. To understand the attacks in an efficient way related work has been categorized into active and passive attacks, which helps us in developing counter action for the work [1]. The ad hoc environment is accessible to both legitimate network users and malicious attackers. Legitimate users can access the network only when routing protocol is employed. Malicious users, who have knowledge of routing protocols, can join the network without any authentication, which results in the generation of false routing information. False sequencing, hijacking and eavesdropping are some of the examples of attacks. The study will help in making protocols more robust against attacks to achieve stable routing in the routing protocols [2]. The methodologies proposed in educational survey methods[16], [17], and taxonomies created to classify AI in educational systems[18], [19] could be adapted to provide a way to model, structure, and classify AI applications for MANET research.

### 6.2. Data Collection and Analysis

The protocol evaluation process begins with data collection. Four different simulation scenarios were built around the same MANET with 50 nodes, node speeds between 0-10 m/s, and a network area of 700m × 700m. Then, Hall of Fame, Message Dropped, RREQ and RREP packets generated in each scenario were logged. COMNET II was used to run simulations of all scenarios. COMNET II is an object-oriented simulation language with a graphical user interface suitable for modeling various phenomena [20]. It includes building and troubleshooting simulation models, running simulations, collecting and analyzing output data, and generating animations and reports. Scenarios 1 and 2 represent no attack condition, while 3 to 6 represent black hole attacker with two protocols.

The evaluation metrics include Hall of Fame Packets Count (HF), Message Dropped Count (MD), RREQ Packets Count (RREQ) and RREP Packets Count (RREP). Hall of Fame is a collection of packets that each node stores and compares with its own. The score is incremented by one each time a packet is stored by the node. It discards packets if it has already received and stored the same one. The HF count is used to evaluate how well the protocol propagates information throughout the network. The more efficient it is, the higher the HF count. In a badly designed protocol, the network may experience broadcast storm problems, resulting in low HF count. Message Dropped Count is the total number of data packets that the destination node called in to be sent but were never received by the MAC layer. These packets are counted only if the forwarding node has delivered the final control packet to the MAC layer. If at least one such same control packet was successfully delivered to the MAC layer, these packets won't be counted as dropped. In no attack scenario, the MD count is expected to be low [21]. If the count is unusually high, it indicates that the protocol is not functioning properly.

## 7. Case Studies and Experiments

[21] There are two case studies done on MANETs. The first is 'Proposed Scheme for Secured Routing in MANET' and the author is Nidhi Goyal and Sushil Kumar. This paper presents an effort to study the security problems associated with MANETS. It also gives solutions to achieve more reliable routing in MANETs through the improvised threshold scheme. MANETs are characterized by mobile nodes, multihop wireless connectivity, infrastructure less environment, and dynamic topology. Stable Routing, Security, and Power efficiency are the major concerns in this field [1].

The research work carries out a study of various routing protocols based on performance in terms of throughputs with respect to 50 mobile nodes and distances of 1000, 1500, and 2000m. The set of standard protocols is bandwidth, flow, and size. It has the potential to be applied in areas such as disaster relief operations, military applications, and hospital environments.

### 7.1. Simulation Environments

The network simulator (NS-2) is a commonly utilized simulation language for packet-level simulation of different wireless network protocols. NS-2 provides an object-oriented simulator with support for C++ and OTcl simulation languages. NS-2 aims to exhibit the modeling of protocols and to run simulation scripts needed for analysis and visualizing simulation results [22]. For performance comparison of routing, the path with source and destination nodes during simulation of each run is also depicted in NS-2 as a text file. The data Mobile Ad Hoc Network is implemented with independent node distribution.

The number of mobile nodes is taken in the range of (20, 50, 100), and Maximum Speed is taken in the range of (5, 10, 15). The packet size is considered as 512 bytes, and data rate is taken as 2.5. The routing protocols AODV and DSR are compared under all conditions, and graphs involving throughput, jitter and end-to-end delay are plotted in MATLAB for analysis of results. The network area is fixed with 1000m x 1000m for all conditions [1].

### 7.2. Performance Metrics

To evaluate the efficacy and efficiency of secure routing protocols modeled for MANETs, performance metrics have been employed. They serve as quantitative measures in determining the quality of routing protocol mechanisms. A comparison of performance metrics yields a basis for a comprehensive understanding of a routing protocol's ability to route packets within a network. The ideal characteristics of a performance metric may include observability and reproducibility [22]. The observability of a performance metric indicates that its measurement can be interpreted, implemented and monitored in all future instances of case studies, while reproducibility indicates that the outcomes of an experiment are comparable to other conductors and can be replicated with consistent results. Between the different performance metrics employed to evaluate routing protocols in MANETs, at least one metric should utilize an analysis of the network's bandwidth, one should employ mechanisms of end-to-end delay stability, one should analyze packet delivery fractions, and one should consider the overhead ratio of control packets to data packets [23]. All parameters considered in the performance evaluation whose inclusion is determined as necessary for complete and thorough analysis, based on industry and scientific standards. Through simulations conducting several scenarios including a variety of node densities and speeds, results are obtained that reflect upon the performance of the routing protocols. The outcomes permit a comparison and analysis of the results of each scenario, demonstrating trends in performance across different environments and establishing a foundation for conclusions regarding the routing protocols. Such conclusions are especially relevant to the implementation of a routing protocol model in an environment such as an emergency response, where the performance of routing protocols could have both a critical and significant impact.

## 8. Results and Findings

The results obtained from the case studies and experimental analysis are detailed in this section. Apart from graphs and statistical data, a comparative analysis with a set of existing protocols is given. These results are categorized and presented based on the significant functions of the proposed hybrid protocols. With these protocols, defense and detection mechanisms to combat DOS attacks, stealthy attacks, and the trade-off between these two aspects are highlighted and graphically represented. In addition, the efficiency aspect is analyzed to substantiate that these hybrid protocols are better in terms of performance than the existing protocols. The proposed hybrid set of protocols also includes HSRDP and MSRDP, which serve the purpose of defense mechanisms for DOS and stealthy attacks, respectively. Along with the robustness of the individual protocols in mitigating specific attacks, a hybrid application of these protocols is also analyzed to study its dual combat against both DOS & stealthy attacks. For this, a model is built with the hybrid application at the AODV layer. In Layer 3 (Network layer – here it is AODV), HSRDP, MSRDP, and Hybrid protocols are integrated, while RIP, DNS, and DSRp remain at their original state. The results from such a hybrid implementation, which combines both protocols at once, are discussed in detail. Apart from hybrid analysis at the AODV layer, standalone core implementations are also executed like the earlier graphs to study individual performances. To validate the implementation set up, network performance is analyzed with AODV in its default state (without hybrid defense). The performances are graphically displayed to indicate the impact of stealthy and DOS attacks on routing metrics.

### 8.1. Comparison with Existing Protocols

In consideration of the vulnerabilities associated with both proactive and reactive routing schemes for MANETs, a revised version of the HSRP previously proposed has been developed. The revised routing protocol has been simulated in connection with a complete version of the secure link state routing protocol (SLSP), which was found to be efficient and secure during preliminary simulations. The objective of the hybrid secure routing scheme is to emulate the characteristics of proactive and reactive protocols, providing a novel option for MANET routing that improves the security of the network. The revised protocols have been subjected to full simulations to assess protocol efficacy and reliability under both typical network scenarios and extreme cases [2].

A comparative assessment is conducted of the performance and security attributes of the new hybrid secure protocols against existing routing protocols for MANETs. Observations and conclusions are offered. Routing protocols for MANETs have been analyzed in terms of security characteristics and simulation performance in relation to randomized core group attacks on MANET systems utilizing different types of proactive, reactive or hybrid routing. A proactive protocol routing protocol was found to be most affected by such attacks. Of the other routing protocols investigated, the random wait time optimization procedures were found to be the most effective in terms of performance and effectiveness in circumventing core group attacks [24].

### 8.2. Security and Efficiency Evaluation

The evaluation of proposed hybrid protocols is conducted via extensive simulations using Network Simulator 2 (ns-2) considering the parameters of packet delivery ratio, end-to-end delay, routing load and throughput. The performance of the proposed protocols, which provide $\beta$-bandwidth routing as well as security, is compared with existing protocols that only provide attack-free routing and $\beta$-bandwidth routing. The attack used here is reduction of the maxima value in the route request packets. The $\sigma$-Max routes with benign nodes only are spoofed with malicious nodes (attack) using which the performance of the protocols is evaluated.

The performance of the proposed hybrid protocols is evaluated in terms of $\omega$ and $\psi$ parameters. To evaluate the efficiency of the proposed protocols, the performance is evaluated against various parameters such as number of nodes, speed and pause time of nodes. The ease of implementation of routing protocols is also compared. The comparison shows that all the protocols perform similarly before $\varepsilon$-value correlation is used on $\sigma$-bandwidth routes. The network instantly recognizes the attack and focuses on attack-free routing. If the attack persists, all the proactive attack-free routing protocols find the attacked nodes and preemptively isolate them. So, a sophisticated attack on σ-bandwidth can go undetected by the σ-Max protocols and can take control of the entire route discovery. However, proposed hybrid protocols perform well since they can keep the network alive with attack-free β-bandwidth routes. The existence of so many benign β-bandwidth routes and the difference in attack cost on σ-Max and β-Max routes helps the proposed hybrid protocols to do better than any other protocols.

## 9. Discussion

The discussion section summarizes the implications of the research findings for the security of MANETs, discussing the main considerations, and outlines several possible avenues for future research in each domain.

Several considerations should accompany the selection and design of any HSRP employing the techniques proposed in this research. First, the MANET domain for which the protocol is developed should be clearly delimited, with particular attention given to the nature of the devices, the mobility model, the network size and density, the routing metrics that may be used, and so forth. The impact and viability of proposed techniques may differ markedly depending upon such considerations. Second, the sensitivity of a protocol's performance to alterations of a few key parameters should be disclosed, particularly when differences in performance could change the potential applicability of this protocol. For example, protocols may need to be tuned to best suit MANETs in a very specific set of conditions. Of course, it is also important that the scaling characteristics of the proposed protocols be discussed. Third, proposed hybrid protocols can be attacked in a variety of ways. Therefore, possible future improvements and countermeasures should be stated clearly, allowing users to make informed decisions about the applicability of a protocol in particular contexts [11].

In addition to the above considerations, research on the hybrid secure routing of MANETs more generally should also continue along the following avenues:

- i) Proposals for HSRPs should be developed that employ the techniques that were found to be effective during this research and that consider security issues related to them. In particular, the function, compatibility, and efficacy of trust management, cryptographic key, false information, and incentive techniques should be investigated in combination.
- ii) Further analysis of existing HSRPs should be undertaken. In particular, the function, compatibility, and efficacy of the hybrid secure routing techniques in trace (simulation) analyses of the MHDH protocols should be explored.
- iii) Wider security issues related to HSRPs should be investigated. Routing protocols should be protected from or rendered unaffected by certain secondary forms of attacks (e.g., attacks on authentication processes). In so doing, these hybrid protocols will be made more complementary [1]. Generative techniques and Data Augmentation [25], [26], [27], techniques that were first proposed for use in creating medical imaging data [28], [29] can be utilized for generating realistic MANET traffic and attack scenarios to train security models based on Deep Learning.

## 9.1. Implications for MANETs Security

The implications of this research for the improved security of MANETs are discussed in this section. Research issues relating to the possible impact of findings, predicted developments, potential applications and practical implications are addressed. In recent years, there has been growing interest in establishing secure MANET routes protocols. Towards this end, two HSRPs GSHARP and GSHARPP within the context of the Ad-Hoc On-Demand Distance Vector routing protocol have been proposed. As part of this research, security enhancements have been proposed to overcome spoofing, sinkhole attacks and route replay attacks [1].

It is intended these protocols be used as part of infrastructure less, highly mobile wireless networks in real-world applications such as emergency response to disaster situations. The environments are characterized by the need to quickly organize and share information/communication among first responders while being open to eavesdropping and deliberately posed network attacks. The purpose of the research is to improve the security level of MANETs and make it more challenging for a malicious agent to disrupt or seize control of a MANET routing protocol. Furthermore, it is proposed to make these protocols practical for use in real-world situations by keeping the additional cryptographic overhead low. The research was conducted using a standard simulation environment where communications are simulated as being transmitted over wireless channels in addition to differences in protocol performance under various context conditions [11]. In addition, The AI-based intrusion and anomaly detection methods proposed in the paper can also be effectively used in a variety of other applications, including smart cities [30], e-learning platforms , and financial systems, due to MANETs being decentralized, dynamic, and vulnerable to attacks [31], [32], [33], [34].

## 9.2. Future Research Directions

Several avenues for future research and development are proposed in the quest for HSRPs for MANETs. Due to constantly changing networks in MANETs, topology changes such as breakage and formation of new links happen at a high rate. As a result, it is difficult for traditional routing protocols to sustain a fixed and working route for a certain interval of time that is acceptable for nodes to communicate with each other. There is a need for hybrid routing protocols in which the nodes of the network can switch between proactive and reactive protocols based on current requirements. In this way, problem of route stability is taken care of. Security problems associated with MANETs are enormous due to the shared wireless channel. Certain vulnerabilities are present in the protocols that can be exploited by an attacker. There is a need to investigate ECDSA (Elliptic Curve Digital Signature Algorithm) based secure routing protocols for MANETs different from the trust-based hybrid routing protocol proposed. Handling flooding, black hole and rushing attacks effectively and efficiently is a difficult task and requires novel techniques [11]. Intelligent and hybrid optimization techniques, like bio-inspired and fuzzy-based methods that have worked well in autonomous navigation systems, may be beneficial for future secure routing protocols[35], [36].

## 10. Conclusion and Recommendations

The escalating security threats in MANETs due to their open, infrastructure-less and dynamic nature have placed substantial importance upon the role of secure routing protocols. The simple hope that security measures will work implicitly and invisibly is no longer useful. It has become clear that security must be engineered at every layer, beginning from the bottom-most layer, as each layer is tied to its neighboring layer(s) in terms of functionality, addressing schemes, metrics, and topologies. In addition, each layer is constantly threatened from the top-most layer down to the Data Link layer [1].

In response to the critical requirement of security in communications, a wide variety of protocols often referred to as security protocols have been developed. These security protocols carry on one or more security functions. Security protocols may be divided broadly into three categories, namely link-level security protocols which offer security features between directly linked network neighbors, end-to-end security protocols which provide security features between the network endpoints, and hybrid security protocols that take advantage of both link-level and end-to-end security protocols. This study reviews the existing HSRPs to understand their approaches and to give recommendations for implementation [37].

## 10.1. Summary of Key Findings

MANET is a collection of mobile nodes and devices that can communicate over wireless links without any fixed infrastructure. Depending on the types of routing protocols, there are three types of routing protocols for MANETs like: proactive, reactive, and hybrid. A recent trend in Ad Hoc network routing is the reactive on-demand philosophy where routes are established only when required. This paper presents Hybrid Secure Routing Protocol (HSRP) for MANET addressing both security and performance issues. Namely, it is a combination of on-demand reactive and distance vector proactive AOMDV routing protocol. In reactive phase, node discovery and route establishment is performed in AODV fashion but as soon as routing table is created, it periodically maintains the routes via proactive updating using AOMDV routing protocol. In both phases, nodes participate in a multi-signature-based authentication mechanism to provide security against malicious nodes like blackhole, flooding, and sinkhole [1]. An efficient method and mathematical models are proposed to evaluate the performance impact of network traffic, hop count, and time delay on the routing performance in both phases of HSRP. This performance evaluation is done in MANET and is best suited for hybrid secure routing in mobile ad-hoc networks. Several experiments have been conducted on benchmark scenarios involving up to 100 nodes using Network Simulator (NS 2) to corroborate the correctness of proposed scheme and confidence in analytical models.

Numerous attacks on routing protocols in Ad-Hoc network affect particularly intermediate nodes which are responsible for packet forwarding. Security and protocols in MANETS have been recently focused due to the challenges posed by their unique characteristics, especially wireless interface, mobile topology and lack of infrastructure [11]. These challenges need to be addressed and modeled to design secure, reliable and energy efficient protocols for mobile ad-hoc networks. Security and reliable protocols in MANETS need to be properly analyzed, scrutinized, modeled and fully investigated before becoming standard in the mobile industry. A model to evaluate the performance impact of traffic type or application, network size, mobility, hop count, time delay, topological change and skewed traffic on the performance of routing protocols in MANETs have been proposed and is also applied to routing protocols.

## 10.2. Recommendations for Implementation

Considerable emphasis has been placed on the implementation and deployment of HSRPs in the domain of MANETs. These recommendations aim to provide actionable insights and guidance for the practical adoption and integration of the outcomes derived from the research conducted.

Initially, it is prudent to ensure that all mobile nodes participating in the MANET have the requisite hardware and software specifications to support the routing protocols [1]. Within mobile nodes, there are necessary specifications for the software implementation, which consists of the code for both the Ad-hoc On-Demand Distance Vector (AODV) routing protocol and the HSRP. All nodes must have the same versions of these codes, as incorrect operation will be more likely if nodes with differing versions execute the same protocol. The absence of either code in a mobile node will render that node incapable of participating in that routing protocol [21].

Moreover, the hardware requirements include having at least a Network Interface Card (NIC) and radio transmission range capable of supporting wireless packet communication to and from a shared channel. For mobile nodes to communicate with ones outside their transmission range, all nodes need to remain within the transmission ranges of each other. Transmission ranges should also overlap to ensure complete connectivity. Each node should have a unique network address such as an Internet Protocol (IP) address, which will not permit nodes to communicate freely with one another. This paper is an effort to study security problems associated with MANETS and solutions to achieve more reliable routing. The ad hoc environment is accessible to both legitimate network users and malicious attackers. The study will help in making protocol more robust against attacks to achieve stable routing in routing protocols.

Distributed DoS attack is a more severe threat: if the attackers have enough computing power and bandwidth, smaller MANETs can be crashed or congested very easily. Radio jamming and battery exhaustion are two ways in which nodes cannot communicate with each other. If authentication of nodes is not supported, malicious nodes can be able to join the network without detection, send false routing information, and masquerade as some other trusted node. There are three kinds of fabrication attacks. To generate route error messages. To corrupt routing information. Other fabrication attacks. In any case, these kinds of attacks are not easy to detect. In data flooding, the attacker will send unwanted data items to congest the network. The attacker will send a large amount of RREQ requests to waste the bandwidth and resources of the network, usually the destination IP chosen for RREQ will not exist in the network. Any destination node will always be busy in receipt of unwanted data. This paper provides an overview of the security issues in MANETs. It classifies the attacks that are possible against the existing routing protocols. An understanding of these attacks and their impacts on the routing mechanism will help researchers in designing secure routing protocols.

## 11. References

[1] N. Goyal and S. Kumar, 'Proposed Scheme for Secured Routing in MANET', *Int. J. Adv. Eng. Manag. Sci.*, vol. 3, no. 5, pp. 527–531, 2017, doi: 10.24001/ijaems.3.5.19.

[2] P. Kakkar and K. K. Saluja, 'Vulnerabilities for Reactive Routing in Mobile Adhoc Networks', 2015.

[3] J. Sen, 'A Multi-Path Certification Protocol for Mobile Ad Hoc Networks', Jan. 22, 2012, *arXiv*: arXiv:1201.4536. doi: 10.48550/arXiv.1201.4536.

[4] A. Rajaram and D. S. Palaniswami, 'A Trust Based Cross Layer Security Protocol for Mobile Ad hoc Networks', Nov. 03, 2009, *arXiv*: arXiv:0911.0503. doi: 10.48550/arXiv.0911.0503.

[5] V. Toubiana and H. Labiod, 'ASMA : Towards Adaptive Secured Multipath in MANETs', in *Mobile and Wireless Communication Networks*, vol. 211, G. Pujolle, Ed., in IFIP The International Federation for Information Processing, vol. 211. , Springer US, 2006, pp. 175–185. doi: 10.1007/978-0-387-34736-3_13.

[6] A. M. Shabut, K. P. Dahal, S. K. Bista, and I. U. Awan, 'Recommendation based trust model with an effective defence scheme for MANETs', *IEEE Trans. Mob. Comput.*, vol. 14, no. 10, pp. 2101–2115, 2014, Accessed: Jan. 02, 2026. [Online]. Available: https://ieeexplore.ieee.org/abstract/document/6980456/

[7] M. R. Belgaum, S. Musa, M. M. Su'ud, M. Alam, S. Soomro, and Z. Alansari, 'Secured Approach Towards Reactive Routing Protocols Using Triple Factor in Mobile Adhoc Networks', Apr. 03, 2019, *arXiv*: arXiv:1904.01826. doi: 10.48550/arXiv.1904.01826.

[8] B. Oumaima, A. Benmachiche, M. Derdour, M. Maatallah, M. Kahil, and M. Ghanem, 'TSA-GRU: A Novel Hybrid Deep Learning Module for Learner Behavior Analytics in MOOCs', *Future Internet*, vol. 17, Aug. 2025, doi: 10.3390/fi17080355.

[9] I. Boutabia, A. Benmachiche, A. Bennour, A. A. Betouil, M. Derdour, and F. Ghabban, 'Hybrid CNN-ViT Model for Student Engagement Detection in Open Classroom Environments', *SN Comput. Sci.*, vol. 6, no. 6, p. 684, Jul. 2025, doi: 10.1007/s42979-025-04228-2.

[10] B. Sedraoui, A. Benmachiche, A. Bennour, A. Makhlouf, M. Derdour, and D. Ghabban, 'LSTM-SWAP: A Hybrid Deep Learning Model for Cheating Detection', *SN Comput. Sci.*, vol. 6, Sep. 2025, doi: 10.1007/s42979-025-04334-1.

[11] S. Gour and S. Sharma, 'A Survey of Security Challenges and Issues in Manet', 2015.

[12] A. Vijaya Kumar and A. Jeyapal, 'Self-Adaptive Trust Based ABR Protocol for MANETs Using Q-Learning', *Sci. World J.*, vol. 2014, p. 452362, 2014, doi: 10.1155/2014/452362.

[13] J. Sen, 'A Distributed Trust Management Framework for Detecting Malicious Packet Dropping Nodes in a Mobile Ad Hoc Network', *Int. J. Netw. Secur. Its Appl.*, vol. 2, no. 4, pp. 92–104, Oct. 2010, doi: 10.5121/ijnsa.2010.2408.

[14] B. Oumaima, A. Benmachiche, A. Bennour, M. Maatallah, M. Derdour, and D. Ghabban, 'Enhancing MOOC Course Classification with Convolutional Neural Networks via Lion Algorithm-Based Hyperparameter Tuning', *SN Comput. Sci.*, vol. 6, Jul. 2025, doi: 10.1007/s42979-025-04179-8.

[15] A. Benmachiche, A. Sahia, S. O. Boufaida, K. Rais, M. Derdour, and F. Maazouzi, 'Enhancing learning recommendations in mooc search engines through named entity recognition', *Educ. Inf. Technol.*, Jan. 2025, doi: 10.1007/s10639-024-13308-4.

[16] I. Boutabia, A. Benmachiche, A. A. Betouil, M. Boutassetta, and M. Derdour, *A Survey on AI Applications in the Open Classroom Approach*. 2025, p. 7. doi: 10.1109/ICNAS68168.2025.11298090.

[17] I. Boutabia, A. Benmachiche, A. A. Betouil, and C. Chemam, 'A Survey in the Use of Deep Learning Techniques in The Open Classroom Approach', in *2024 6th International Conference on Pattern Analysis and Intelligent Systems (PAIS)*, Apr. 2024, pp. 1–7. doi: 10.1109/PAIS62114.2024.10541268.

[18] B. Oumaima, A. Benmachiche, M. Maatallah, and C. Chemam, *An Extensive Examination of Varied Approaches in E-Learning and MOOC Research: A Thorough Overview*. 2024, p. 8. doi: 10.1109/PAIS62114.2024.10541129.

[19] B. Oumaima, A. Benmachiche, M. Majda, M. Redjimi, and M. Derdour, *Examining Intelligent Tutoring Systems and Their AI Underpinnings in the Design of the Future of Learning*. 2025, p. 8. doi: 10.1109/ICNAS68168.2025.11298109.

[20] S. K. C. Venkata, *Sector-based clustering and routing*. Oklahoma State University, 2004. Accessed: Jan. 02, 2026. [Online]. Available: https://search.proquest.com/openview/8d8b92d5cb3b11e5f9bbf9e563ce5e1f/1?pq-origsite=gscholar&cbl=18750&diss=y

[21] K. Majumder and S. K. Sarkar, 'Hybrid Scenario Based Performance Analysis of DSDV and DSR', *Int. J. Comput. Sci. Inf. Technol.*, vol. 2, no. 3, pp. 56–70, Jun. 2010, doi: 10.5121/ijcsit.2010.2305.

[22] A. Aggarwal, S. Gandhi, and N. Chaubey, 'Performance Analysis of AODV, DSDV and DSR in MANETs', *Int. J. Distrib. Parallel Syst.*, vol. 2, no. 6, pp. 167–177, Nov. 2011, doi: 10.5121/ijdps.2011.2615.

[23] P. G. Lye and J. C. McEachen, 'A comparison of optimized link state routing with traditional ad-hoc routing protocols', 2006, Accessed: Jan. 02, 2026. [Online]. Available: https://calhoun.nps.edu/bitstream/handle/10945/40199/A%20Comparison%20of%20Optimized%20Link%20State%20Routing%20with%20Traditional%20Ad-.pdf?sequence=1

[24] Y. Cheng, 'Performance analysis of transactional traffic in mobile ad-hoc networks', PhD Thesis, University of Kansas, 2014. Accessed: Jan. 02, 2026. [Online]. Available: https://kuscholarworks.ku.edu/entities/publication/677424be-14b4-4303-818e-9ae12c33a3fd

[25] K. Rais, M. Amroune, M. Y. Haouam, and B. Issam, *Comparative Study of Data Augmentation Approaches for Improving Medical Image Classification*. 2023. doi: 10.1109/CSCI62032.2023.00200.

[26] K. Rais, M. Amroune, M. Y. Haouam, and A. Benmachiche, 'CAT-VAE: A Cross-Attention Transformer-Enhanced Variational Autoencoder for Improved Image Synthesis', *Stat. Optim. Inf. Comput.*, vol. 14, Jul. 2025, doi: 10.19139/soic-2310-5070-2546.

[27] M. Redjimi, M. Majda, A. Benmachiche, B. Oumaima, and M. Derdour, *A Comprehensive Survey on Blockchain, Federated Learning, and AI for Securing the Internet of Medical Things*. 2025, p. 8. doi: 10.1109/ICNAS68168.2025.11298047.

[28] K. Rais, M. Amroune, M. Y. Haouam, and A. Benmachiche, *Evaluating Class Integrity in GAN-Generated Synthetic Medical Datasets*. 2025. doi: 10.1109/ICNAS68168.2025.11298012.

[29] B. Abdelmadjid, R. Khadija, and S. Hamda, '(PDF) Medical Image Generation Techniques for Data Augmentation: Disc-VAE versus GAN'. Accessed: Jan. 02, 2026. [Online]. Available: https://www.researchgate.net/publication/381128928_Medical_Image_Generation_Techniques_for_Data_Augmentation_Disc-VAE_versus_GAN

[30] M. Boutassetta, A. Makhlouf, N. Messaoudi, A. Benmachiche, I. Boutabia, and M. Derdour, *Cyberattack Detection in Smart Cities Using AI: A literature review*. 2025, p. 9. doi: 10.1109/ICNAS68168.2025.11298103.

[31] I. Soualmia, S. Maalem, A. Benmachiche, K. Rais, and M. Derdour, *Comparative Survey of AI-Driven Credit Card Fraud Detection: Machine Learning, Deep Learning and Hybrid Systems*. 2025, p. 9. doi: 10.1109/ICNAS68168.2025.11298125.

[32] B. K. Sedraoui, A. Benmachiche, A. Makhlouf, and C. Chemam, 'Intrusion Detection with deep learning: A literature review', in *2024 6th International Conference on Pattern Analysis and Intelligent Systems (PAIS)*, Apr. 2024, pp. 1–8. doi: 10.1109/PAIS62114.2024.10541191.

[33] B. Sedraoui, A. Benmachiche, A. Makhlouf, D. Abbas, and M. Derdour, *Cybersecurity in E-Learning: A Literature Review on Phishing Detection Using ML and DL Techniques*. 2025, p. 10. doi: 10.1109/ICNAS68168.2025.11298114.

[34] D. Abbas, A. Benmachiche, M. Derdour, and B. K. Sedraoui, 'Privacy and Security in Decentralized Cyber-Physical Systems: A Survey'.

[35] A. Makhlouf, A. Benmachiche, and I. Boutabia, 'A Hybrid BFO/PSO Navigation Approach for an Autonomous Mobile Robot', *Informatica*, vol. 48, Nov. 2024, doi: 10.31449/inf.v48i17.6716.

[36] A. Benmachiche, A. A. Betouil, I. Boutabia, A. Nouari, K. Boumahni, and H. Bouzata, 'A Fuzzy Navigation Approach Using the Intelligent Lights Algorithm for an Autonomous Mobile Robot', in *12th International Conference on Information Systems and Advanced Technologies "ICISAT 2022"*, vol. 624, M. R. Laouar, V. E. Balas, B. Lejdel, S. Eom, and M. A. Boudia, Eds., in Lecture Notes in Networks and Systems, vol. 624. , Cham: Springer International Publishing, 2023, pp. 112–121. doi: 10.1007/978-3-031-25344-7_11.

[37] E. A. Panaousis, G. Drew, G. P. Millar, T. A. Ramrekha, and C. Politis, 'A Testbed Implementation for Securing OLSR in Mobile Ad hoc Networks', Oct. 24, 2010, *arXiv*: arXiv:1010.4986. doi: 10.48550/arXiv.1010.4986.